AIMC 2023

# Are words enough?

**Jorge Forero**[1] **Gilberto Bernardes**[2] **Mónica Mendes**[1]

[1]ITI LARSyS, [2]INESC TEC

# ARE WORDS ENOUGH?

**On the semantic conditioning of affective music generation.**

Music has been commonly recognized as a means of expressing emotions. In this sense, an intense debate emerges from the need to verbalize musical emotions. This concern seems highly relevant today, considering the exponential growth of natural language processing using deep learning models where it is possible to prompt semantic propositions to generate music automatically.

This scoping review aims to analyze and discuss the possibilities of music generation conditioned by emotions. To address this topic, we propose a historical perspective that encompasses the different disciplines and methods contributing to this topic. In detail, we review two main paradigms adopted in automatic music generation: rules-based and machine-learning models. Of note are the deep learning architectures that aim to generate high-fidelity music from textual descriptions. These models raise fundamental questions about the expressivity of music, including whether emotions can be represented with words or expressed through them.

We conclude that overcoming the limitation and ambiguity of language to express emotions through music, some of the use of deep learning with natural language has the potential to impact the creative industries by providing powerful tools to prompt and generate new musical works.

## I. INTRODUCTION

Juslin & Västfjäll (2008) describe emotions as part of the broader domain of affects, which encompasses moods and personal preferences. Emotions and moods are distinguished based on their brevity and intensity (Scherer, 2000). Historically, music is regarded as expressing[1] emotions (Davies, 1994; Juslin & Laukka, 2003). However, evoking emotions from music is a complex phenomenon studied from different methodological perspectives in psychology, cognition, music theory, and music information retrieval, among other areas.

An intense debate emerges from the need to verbalize musical emotions. Eerola and Vuoskoski (2013) pinpointed fundamental questions: what are the putative emotions induced by music, and which components contribute to these? Are the processes involved learned or universal? This matter seems relevant today with the exponential growth of natural language processing, using deep learning models to generate high-fidelity music from text descriptions. However, affective music generation conditioned by semantics is a challenging task that tackles several problems, such as the ambiguity of language, the subjectivity of emotions, and the complexity of the music structure.

This paper exposes a historical perspective of automatic music generation from rules-based expert systems to deep learning models. We aim to identify and discuss the limits of expressing affective music conditioned by semantics. We conclude that although there are potential downsides to using these tools, their benefits in inspiration and efficiency should not be ignored.

The remainder of this paper is structured as follows. **Section II** presents the musical context of emotions by proposing a historical overview of categorical and dimensional models of emotion. **Section III** reviews how low- and mid-level music structural features identified in the literature relate to emotions. **Section IV** delves into the affective algorithmic composition (AAC), dividing models into rules-based and data-driven methods to create music capable of evoking a specific emotional state. Finally, **Section V** discusses the limits of affective musical generation conditioned by emotions.

## II. THE MUSICAL CONTEXT

Models for analyzing and measuring emotions can be grouped into categorical, dimensional, and prototype-based models. Categorical models suggest that emotional experiences can be represented by discrete adjectives such as happiness, sadness, anger, surprise, disgust, and interest (Ekman, 1984-1992; Izard, 1977; Panksepp, 1982). In contrast, dimensional models conceptualize emotions based on their position along affective dimensions, such as arousal and valence (Wundt, 1905; Russell, 1980; Thayer, 1989). Prototype theories propose that language shapes how we conceive emotions and that emotions can be categorized based on their similarity to prototype exemplars (Shaver et al., 1987). Next, we detail the two first approaches due to their relevance to affective music generation.

### CATEGORICAL MODELS IN MUSIC

The verbalization of emotions conveyed through music has been a topic of interest and intense philosophical discussion traced back to at least since ancient Greece (Budd, 1985). However, the cornerstones for systematically exploring the interplay between emotions and music only appeared in the late-19th century.

Gilman (1892) explored the power of music to awaken definite ideas and emotions in the listener. He invited a group of 28 "music-interested" evaluators to listen to different fragments of Western classical music in an experimental concert. The researcher posed 11 open-ended questions or prompts to the subjects, which were to be reported in a notebook containing instructions. Downey (1897) conducted another experimental concert where 22 evaluators of varying musical proficiency freely reported their impressions of a live performance without prior questions or prompts. The concert involved the interpretation of six piano works of western classical music. He concluded that superimposing emotional states require music with high expressiveness.

Hevner's (1935) work is widely recognized as an early systematic approach to the categorical classification of emotions in music. She proposes a taxonomy of eight categories resulting from grouping 66 adjectives (see Media1). These semantic categories are arranged in 1936 into a circle to conceptualize their relationship. This spatial configuration is known as Hevner's adjective circle. Farnsworth (1954) and Asmus (1985) have followed Hevner's model. Schubert (2003) also revised Hevner's model by asking 133 experienced musicians to assess a list of 91 adjectives in describing the emotional expressiveness of music. Juslin and Laukka (2003) grouped various authors' adjectives up to that time into five basic categories (anger, fear, happiness, sadness, and love-tenderness).

> Visit the web version of this article to view interactive content.

Media1: Hevner adjectives list.

Since 2007, the Music Information Retrieval Evaluation eXchange (MIREX) has employed a categorical model in the MIREX Audio Mood Classification tasks (Hu et al., 2008). This model categorizes emotions into five discrete clusters, each comprising five to seven related emotions (adjectives). Lauriere et al. (2009) construct a semantic mood space from last.fm tags using Latent Semantic Analysis. They compare this community-based semantic space with expert representations from Hevner and the clusters derived from the MIREX Audio Mood Classification task.

Zentner et al. (2008) developed a scale specific to the music domain called the Geneva Emotional Music Scale (GEMS). To create this scale, the researchers conducted four studies. The first and second studies aimed to compile a list of terms for perceived and induced emotions. One of the shorter versions is the three-term scale, which includes unease, vitality, and sublimity (see Media2). Coutinho and Scherer (2017) updated the GEMS scale to create the Geneva Music-Induced Affect Checklist (GEMIAC).

> Visit the web version of this article to view interactive content.

Media2: GEMS Emotional responses to music.

## DIMENSIONAL MODELS IN MUSIC

The Semantic Differential (SD) technique was created by Osgood (1953) and further developed by Osgood et al. (1957) as a tool to evaluate meaning. Initially, the scale was semantically oriented through polarized and/or graduated verbal-quantitative descriptions. Specifically, research has identified that affective judgments on bipolar semantic differential scales tend to converge into three dimensions. Some authors argue that the two-dimensional arousal-valence model may not adequately represent music's expressive possibilities (Collier, 2007; Ilie & Thompson, 2006) and opt for models with a third dimension, such as power, intensity, and dominance. Additionally, dimensions that appear more common in music, such as tension-release, crescendo-diminuendo, and accelerando-ritardando, are also considered (Eerola et al. 2009).

A dimensional approach to emotion may be particularly well-suited for examining shifts or conflicts in emotional expression. The use of dimensional models is spatially applicable when physiological measures of emotional response to induce music are required from listeners. In the 21st century, research has increasingly investigated the relationship between music and emotion in neuropsychology (Peretz, 2010). Furthermore, listening to music activates various brain regions involved in emotion and reward processing (Koelsch, 2020).

## III. MAPPING EMOTIONS TO MUSICAL STRUCTURES

Once the emotional framework is established, researchers can explore how different structural music elements relate to the emotional experience. In this context, studies have consistently demonstrated that various musical properties, including rhythm, melody, harmony, and timbre, play a significant role in shaping our emotional response to music. In this section, we detail the intricate relationship between these properties and emotions, pinpointing the different combinations of musical features that elicit a range of emotional experiences.

### RHYTHM AND EMOTIONS

The tempo of the music has been consistently linked to the emotional response it evokes. Faster tempos are generally associated with more positive valence and higher arousal than slower tempos. Rigg (1940) used six tempos and five musical samples, with participants evaluating the music using a bipolar semantic differential scale with serious-sad / pleasant-happy criteria. After this initial judgment, auditors were to attempt a finer discrimination by classifying the phrase under eleven emotional categories. The study's results revealed that faster tempos tend to make the music happier, while slower tempos tend to make it sadder.

Some studies have refuted this causal relationship. Holbrook and Anand's (1990) perceptually assessed 14 variations of the same jazz song with different tempos and found a non-linear relationship between tempo and emotional response, with the peak emotional response occurring at 108 bpm. Iwanaga and Tsukamoto (1998) investigated tempo preferences using known and unknown melodies with different tempos. Fast tempos were rated highly active, regardless of the tempo or prior listening experience. On the other hand, affects, including preferences, demonstrated an inverted U-shaped relationship with tempo variation. Specifically, the most preferred tempo was the designated tempo for known melodies, while moderate tempos ranging from 109 to 130 beats per minute (BPM) were preferred for unknown melodies.

Kamenetsky et al. (1997) demonstrated that variations in dynamics resulted in higher ratings of likability and emotional expressiveness than those for tempo. The study included four conditions: unvarying tempo and dynamics, variations in tempo only, variations in dynamics only, and variations in tempo and dynamics. Gabrielsson's (1988) study explored the relationship between tempo and note density, demonstrating that, in certain instances, these two features may have an additive effect on emotional activation. Specifically, faster tempos with higher note density increased activation, while slower tempos with lower note density led to reduced activation (Madison and Paulin 2010).

Van der Zwaag et al. (2011) investigated the influence of the musical characteristics of tempo and percussiveness on emotions. Their results show that increased tempo increased reported arousal and tension and decreased heart rate variability. Also, the level and frequency of skin conductance responses (SCR) were reported to increase with percussive sounds. Conversely, Brownlow (2017) discovered that the SCR was significantly higher in fast-tempo music. These results suggest that the faster the tempo, the greater the physiological and emotional activation. However, Bramley et al. (2016) found that tempo alone is not the determining factor in physiological arousal.

## MELODY AND HARMONY: THE (TONAL) TENSION

There is no trivial answer to why minor tones tend to be associated with negative valences and if this proposition can be proved as truth. As described below, systematic attempts approach this question by measuring the tension these musical features evoke.

Heinlein (1928) investigated the possible emotional effects of music by varying tonality. To do this, he asked listeners to respond to self-report their experiences after listening to the chords recorded on a Duo-Art piano roll. The author did not note significant differences and concluded that the fixity of feeling-tone seems to be a matter of training. However, Hevner (1935) criticized Heinlein's research methodology, arguing that the chosen chords were outside their musical context and thus could not be accurately measured. To overcome this issue, Hevner proposed conducting experiments using short musical fragments from western classical music, with complete musical ideas and two identical versions in rhythm, tempo, and intensity but transposed to transform their tonality.  The study found that listeners consistently assigned positive emotional labels to music in major keys and negative labels to music in minor keys, regardless of their musical training.

Gerardi and Gerken (1995) manipulated a pre-existing melody to generate four stimuli. The experiment changed the piece's mode (from major to minor or vice versa) and the melodic contour (up to down or down to up). The results showed that the modality of the melody affected valence perceived by persons of eight years old or older participants. The majority of happy responses were given for major mode melodies. The effect of melodic contour was slightly weaker, with a trend for happy melodies to be associated with rising contours and sad melodies with descending contours. Gregory et al. (1996) conducted a study to examine the development of emotional responses to music in young children. They used modified and original versions of eight nursery tunes and manipulated two parameters: modality (major/minor) and harmony (absence/ presence). Their results showed that a positive relationship between modality and valence was established by the age of seven to eight years.

Narmour's (1992) Implication-Realization (I-R) Model is a detailed formalization of Meyer's work on musical expectation (Meyer, 1956). Margulis (2004) developed a model that expands on the Implication-Realization model by incorporating a melodic attraction factor and local and global expectations. Her model assigns expectedness ratings to melodic events and associates them with three types of tension - surprise-tension, denial-tension, and expectancy-tension.  Farbood (2012) suggested that although tension and arousal are not synonymous, they are closely interrelated. Broadly, a rise in tension has been equated with an increase in excitement or intensity. In contrast, a decrease in tension has been described as a feeling of relaxation or resolution (Farbood 2017).

## CORRELATIONS BETWEEN RHYTHM AND HARMONY

Erich Sorantin (1932) delved into the issue of musical expression. Sorantin's investigations resulted in two primary categories of musical figures: rhythmical and melodic. Rhythmical figures represent the interaction of auditory and visual senses and depict diverse types of motion, such as swinging, trembling, and rotation. On the other hand, melodic figures can be classified by the author into two figure categories: lamentation and joy. Figures of lamentation were recognized by descending melodic lines in minor seconds, trochaic rhythmic patterns, dissonances on culminating points, and slow tempo with restricted melodic movement. In contrast, figures of joy are characterized by a symmetrical and balanced melodic line, the use of the interval of the fourth, simplicity of harmony in the major mode, the absence of dissonances, predominant rhythmic emphasis through accentuation and staccato, and frequent occurrence of trills.

Schellenberg et al. (2000) determined that pitch is more significant than tempo or rhythm in conveying emotions. The study's findings indicate that rhythm impacts music's emotional rating only when pitch changes accompany it. Furthermore, the study found that differences in pitch had a more significant influence on the ratings of different melodies than differences in rhythm. Whenever rhythm affected the ratings, it was always in combination with pitch. Collier and Hubbard (2001) confirmed these results in their studies using scales as stimuli, demonstrating that pitch height and contour (ascending and descending C major scale) may be more crucial than the mode to express happiness. Ascending tones were rated as happier, brighter, faster, and speeding up more than descending tones.

Husain and collegues (2002) found that the subjective enjoyment rating for the combination of major key and fast tempo (165 bpm) is significantly higher than that for combinations of minor key and slow tempo (60 bpm). Similarly, Gagnon and Peretz (2003) presented a stimulus comprising a melody with ten notes, varying the tonality (major, minor) and tempo (110, 165, 220 bpm). The combination of major keys and fast tempos was rated more positively (happy) than its minor and slow counterpart, which was associated with sadness. The authors further concluded that tempo has a more pronounced effect on emotional valence than tonality. Webster and Weir (2005) further expanded upon this idea by utilizing short five-measure sequences comprising 12 possible combinations: 2 textures (harmonized/non-harmonized) x 2 modes (major/minor) x 3 tempos (72, 108, and 144 bpm), which were evaluated on a sad/happy bipolar scale. The study concluded that as tempo increases, the happiness rating also rises linearly for non-harmonized samples in major tonalities but not linear or quadratic for harmonic samples in minor ones.

Swaminathan and Schellenberg (2015) provided a summary of previous research on the emotional effects of music. They suggest that certain combinations of musical elements can evoke complex or ambiguous emotions in listeners. For example, fast-tempo music is typically associated with happier emotions, while minor modes are associated with sadness. By varying tempo and mode independently, it is possible to create music with conflicting structures that elicit mixed emotions in listeners. [Gabrielsson (2016)](#) concludes that the major mode is not a prerequisite for the perception of happiness. For instance, music in the minor mode, played at a fast tempo, can also convey a happy emotional state.

FURTHER CORRELATIONS: TIMBRE & GENRE

[John Sloboda (1991)](#) developed an experiment where participants were instructed to select their preferred songs and indicate specific sections of these pieces that elicited such responses as crying, shivering, or laughter. Furthermore, [Behrens and Green (1993)](#) investigated the capacity to recognize emotional expression in improvised music. The study required each performer to convey the emotions of sadness, anger, and fear, one per improvisation. The results indicated that the subjects were relatively accurate in assessing the emotional content of the improvisations. Nevertheless, the participants' accuracy depended on either their level of musical experience, the instrument used by the performer, and the emotion expressed.

[Paraskeva and McAdams (1997)](#) demonstrated that high-level attributes such as orchestration could impact the perception of tension in the music of various genres. Past research has also explored the effects of timbral components on tension perception, including roughness, brightness, and density ([Plomp & Levelt, 1965](#); [Krumhansl, 1996](#); [Pressnitzer et al., 2000](#)). The pitch register can be classified under the category of timbre change. [Granot and Eitan (2011)](#) reported no significant variations in responses to dynamics and tempo between musicians and non-musicians. The authors also established that the most influential factor was dynamics, followed by pitch register.

[Livingstone et (2010) al.](#) presented CMERS, a Computational Music Emotion Rule System for the real-time control of musical emotion that modifies features at both the score level and the performance level (see Image 1).

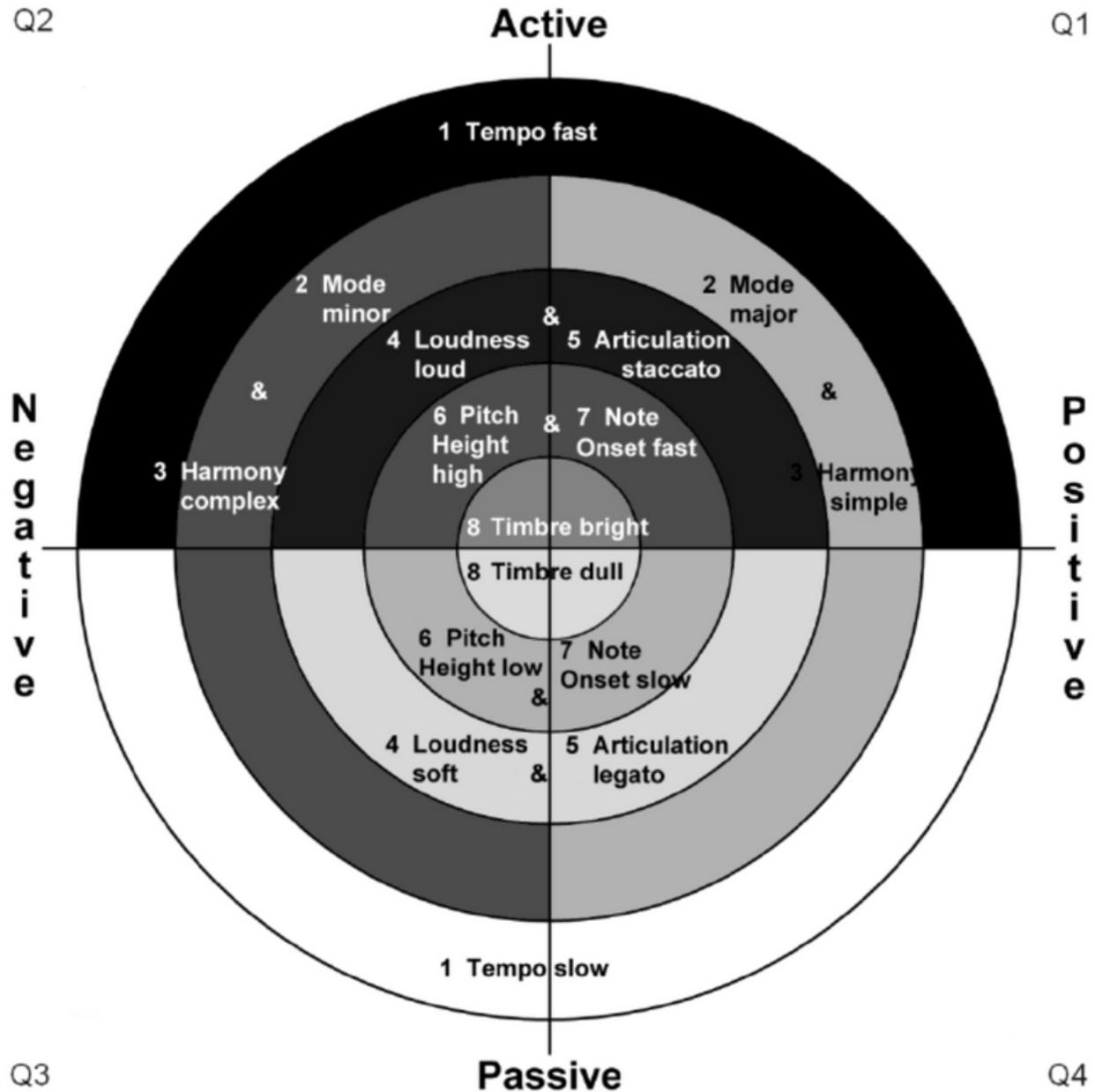

Image 1: Set of Primary Music-Emotion Rules mapped on to the 2DES. Adapted from Livingstone and Thompson (2006).

Panda et al. (2020) present a survey on the existing emotionally-relevant computational audio features, supported by the music psychology literature on the relations between eight musical dimensions (melody, harmony, rhythm, dynamics, tone color, expressivity, texture, and form) and specific emotions.

Visit the web version of this article to view interactive content.

2D Affective Space Rules

## IV. AFFECTIVE ALGORITHMIC COMPOSITION

Affective Algorithmic Composition (AAC) focuses on developing methods for conditioning the emotional content of music composed by algorithmic means to evoke a desired emotional response in the listener. Commonly used methods are rule-based and machine-learning systems such as neural network architectures (Dash, 2023).

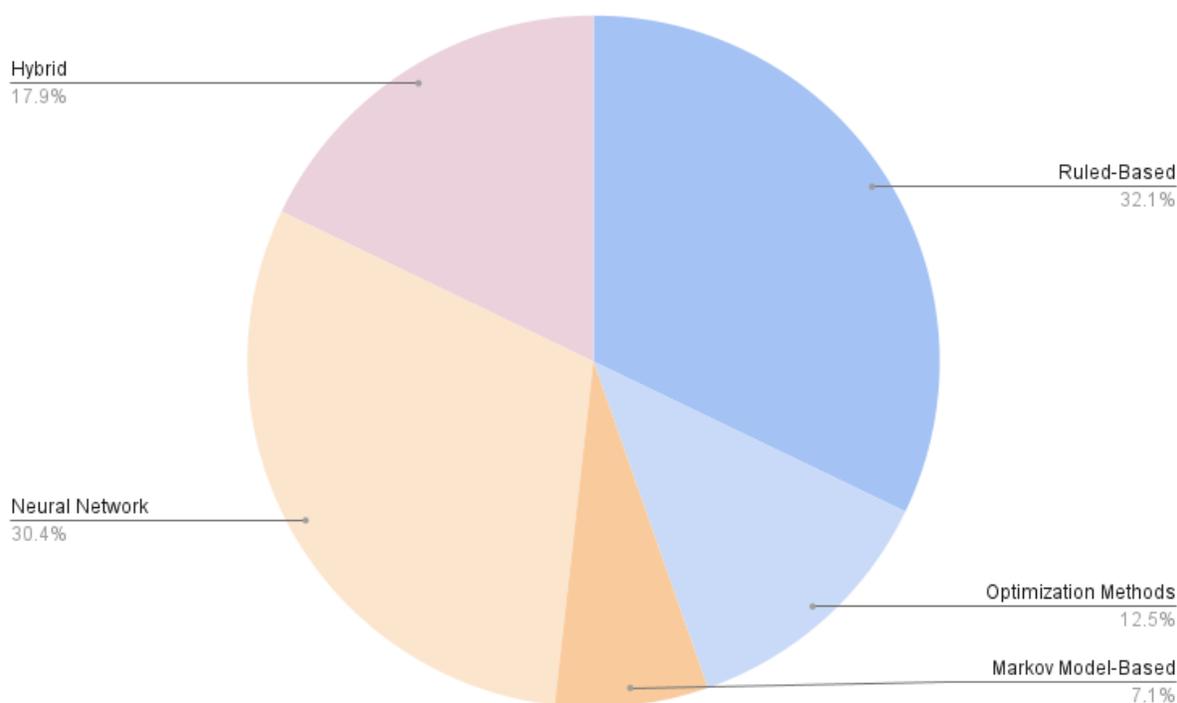

Image 2: Common methods used for AAC.
Based on **Dash, 2023** (N=56)

### RULE-BASED METHODS

Rule-based methods are employed to define the musical rules of the compositions. These musical rules are represented in the form of either mathematical equations or logical statements.

TransProse is a rule-based system for composing piano melodies (Davis & Mohammad, 2014). The system divides a given text into sections and uses a grammar-based approach to assign an emotional label to each section. TransProse employs pre-defined rules based on music theory to compose a melody for each section, controlling factors such as scale, tempo, octave, and notes. Scirea et al. (2017) proposed a real-time framework called MetaCompose for generating background music for video games. The MetaCompose framework uses an expert system called *Real-Time Affective Music Composer,* controlling four musical attributes: volume, timbre, rhythm, and dissonance. Forero et al. (2022) developed a method to create affective virtual environments adopting speech as the leading interaction. They used two machine learning models to predict emotions from speech, which were mapped to four attributes for controlling tonal tension: dissonance, distance to key, tempo, and rhythm.

**OPTIMIZATION METHODS**

Optimization Methods (OM) are rule-based expert systems because they must define a fitness function. Evolutionary algorithms (EAs) applied to affective generative music apply OM that uses a fitness function to evolve an original piece to a specific style. Defining this function is challenging; thus, most EAs employ an interactive method, where human subjects assess whether the generated pieces correspond to a target emotion. Kim and Andre (2004) proposed a perceptual interface with genetic optimization methods to adapt to user preferences. The authors employed two approaches to evaluate the fitness of a rhythm generator. The first method relies on explicit user judgments and is used for new users to train the system. For users the system knows already, the fitness is computed based on their physiological response. Zhu et al. (2008) proposed an approach to generate emotional music using interactive genetic algorithms based on the KTH rule system[2]. Xu et al. (2010) proposed an emotional harmony composition system that incorporates two distinct types of rules: harmony composition rules, which are employed to compose harmony, and harmony emotion rules, which indicate the mapping between harmonic elements and a specific emotion, either happiness or sadness. Fukumoto and Nomura (2018) proposed a Distributed Interactive Genetic Algorithm (DIGA) for generating four-bar piano melodies with controllable brightness. Rocha de Azevedo et al. (2021) experimented with generating piano music using template pieces in MIDI format as a mood guide. The authors randomly generated a population of pieces with two contrasting moods: happy and sad.

## DATA-DRIVEN METHODS: THE MACHINE LEARNING ERA

Data-driven automated music generation (AMG) techniques aim to identify and learn from a dataset's inherent music features. Next, we detail different models and architectures using machine learning for music, such as Markov model-based approaches and neural network-based systems (recurrent neural network architectures and variational autoencoders).

**MARKOV MODEL-BASED APPROACHES**

The parameters of a Markov model can be adjusted using rule-based or data-driven methods. Amani Indunil Soysa and Lokuge (2010) proposed a solution for harmonization by introducing an interactive framework called "ChordATune." The framework generates chord progressions and harmonizes melodies based on user emotions. The melody has to be inputted into the system as an audio file, and a pitch class profile is created at runtime to represent the pitch content of the file over time. The authors employed a machine learning approach using Hidden Markov Model (HMM) and dynamic programming to generate chord progressions for an input melody while incorporating the user's emotional factor. Monteith et al. (2012) introduce a system that uses statistical models derived from a corpus of songs. The system composes melodies using n-gram models that represent pitch intervals frequently found in the training corpus for the desired emotion. In addition, Hidden Markov Models are also employed to generate harmonies similar to those present on a particular corpus. Maniktala et al. (2020) presented MINUET (Mood Into Note Using Extracted Text), a semantic system that uses sentiment analysis to generate music for textual narrative segments procedurally.

### NEURAL NETWORK-BASED SYSTEMS

Various neural network-based architectures have been designed for generating affective music. Most of these systems employ either a Recurrent Neural Network (RNN), a Long Short-Term Memory (LSTM), or a Variational Auto Encoder-Generative Adversarial Network (VAE-GAN) as the fundamental architecture of the system.

Daly and colleagues (2015) propose an AAC system for generating a stimulus set that covers nine discrete sectors of a two-dimensional emotion space, using a 16-channel feed-forward artificial neural network. The system generates a set of short music pieces rendered using a sampled piano timbre.

### RECURRENT NEURAL NETWORKS (RNNs)

Recurrent neural networks (RNNs) were developed to handle sequences and time series data. Using their internal state, RNNs can process input sequences of variable length and thus have become the most popular choice for music generation.

Madhok et al. (2018) proposed a framework named SentiMozart to generate music based on the predicted emotion of users using a two-model approach. The first model, the Face Classification Model, identifies and classifies facial expressions into one of the seven major emotional categories using Convolutional Neural Networks (CNNs). The second model, the Music Generation Model, is a Doubly Stacked LSTM architecture that generates music corresponding to the predicted emotion.

Ferreira and Whitehead (2019) introduced a novel approach to music generation and emotion analysis using a multiplicative long short-term memory (mLSTM) based model. Ferreira et al. (2020) introduced Bardo Composer, a system that generates music for tabletop role-playing games based on players' moods.

Zheng et al. (2021) presented EmotionBox. The model is trained on a recurrent gated unit (GRU) using the piano MIDI dataset labeled by Zhao et al. (2019). EmotionBox can be controlled by a temperature hyperparameter, and it generates music by modifying pitch histogram and note density, respectively, representing mode and tempo.

**VARIATIONAL AUTOENCODERS**

Another widely used neural network architecture is variational autoencoders (VAE). One notable example is Google's MusicVAE (Roberts et al., 2018), which employs a hierarchical VAE to learn long-term structures. Tan and Herremans (2020) proposed a framework named Music FaderNets that can learn high-level feature representations with limited data by modeling their corresponding quantifiable low-level attributes. Using arousal as an example of a high-level feature, the authors showed that the model successfully learns the intrinsic relationship between arousal and its corresponding low-level attributes (rhythm and note density) with only 1% of the training set labeled. Finally, using the learned high-level feature representations, the authors explored the application of their framework in style transfer tasks across different arousal states.

Tiraboschi et al. (2021) proposed a set of techniques for real-time music generation applications that employ electroencephalogram (EEG) sensor data. The authors presented a Probabilistic Graphical Model to establish a mapping from the valence-arousal space to the latent variables of MusicVAE. This approach enables the model to capture the relationship between the user's emotional state, as reflected in the EEG data, and the generated music, which can be adjusted in real-time to match the user's emotional state. VAE-GAN networks comprise a sequence-to-sequence architecture that connects an encoder and a decoder/generator in a series. Huang et al. (2020) proposed an AMG System using CVAE-GAN. The system uses a music database with emotional tags as input, and deep learning trains the CVAE-GAN model as the framework to generate music segments corresponding to specific emotions.

**SEMANTIC CONDITIONED NEURO AUDIO GENERATION.**

Recent advancements in text-to-image generation have inspired researchers to explore generating audio from high-level, sequence-wide captions (Yang et al., 2022; Kreuk et al., 2022), similar to text-based image generation.

Huang and colleagues (2022) developed MuLan, a novel acoustic model that links music audio directly to unconstrained natural language music descriptions. This model uses a "two-tower, joint audio-text embedding mode" and was trained using 44 million music recordings and weakly-associated, free-form text annotations (see Image 3).

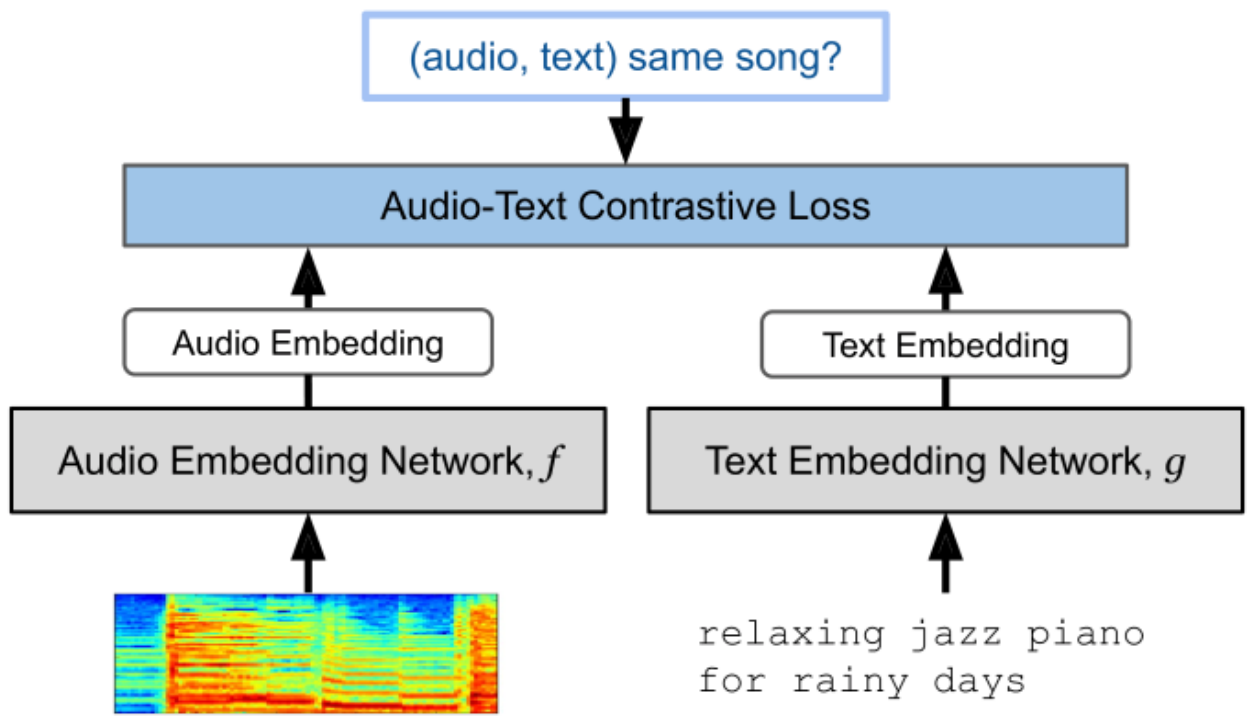

Image 3: MuLan *Learning framework diagram.*

Agostinelli and colleagues (2022) introduced AudioLM as a framework for generating high-fidelity audio conditioned by semantics. This model can generate realistic speech or piano music with long-term coherence lasting several seconds through unsupervised training on audio-only corpora without requiring any annotations. Building upon AudioLM and MuLan, the MusicLM model was developed for generating high-quality music from text descriptions using multi-stage autoregressive modeling. When trained on a large dataset of unlabeled music, MusicLM can generate coherent and long musical pieces at 24 kHz, even for text descriptions of significant complexity. To address the sparse data available for this task, the authors introduced MusicCaps, a high-quality music caption dataset containing 5.5k examples prepared by expert musicians. Overall, MusicLM outperforms previous systems, such as Mubert[3] and Riffusion (Forsgren & Martiros, 2022), regarding quality and adherence to the caption.

## V. DISCUSSION

An intense debate emerges from the need to verbalize musical emotions. Systems of representation are fundamental pieces that allow us to measure and interpret the data. In this scope, we have presented the two complementary perspectives adopted in automatic music generation: rules-based approaches and machine learning models. Within the latter models, deep learning seems to push humans to the end of the process, writting text to produce music. However, automatic music composition conditioned by emotions is a challenging task that involves several problems, such as:

1. *The ambiguity of language*. Language is inherently ambiguous and subjective, leading to different musical interpretations and making it challenging to capture the intended emotions in the music structure.
2. *The subjectivity of emotions*. The experience of emotions is unique to each individual. Therefore, affective music generation models mainly capture the intended emotions for some listenersly.
3. *The complexity of music*. Music is a complex art form that involves multiple components, including melody, harmony, rhythm, and timbre. Generating musically coherent and emotionally expressive music is a challenging task that requires models to capture the relationships between these components and emotions, which have been shown to be non-linear.
4. *Evaluation of generated music*. Evaluating the quality of generated music is a subjective task that personal preferences and biases can influence. Therefore, it can be challenging to objectively evaluate the effectiveness of models that attempt to generate music conditioned by language and emotions.

Furthermore, we should consider some techno-political issues, such as the lack of public large-scale datasets and policies that avoid information monopolies. Another concern is the need for more transparency in deep learning algorithms. Unlike traditional rule-based systems, deep learning algorithms operate as "black boxes" that make decisions based on complex mathematical models that are difficult to interpret. This lack of transparency makes it difficult to understand how these algorithms make decisions and to ensure that they are making decisions that are ethical and aligned with human values.

Still, we should consider that even in a well-defined frame of reference, with the best possible categorization for an individual, the model is always a reduction that considers an average. And that is a huge problem that could lead us to an undesirable convergence and oversimplification of music.

On the other hand, one of the main advantages of deep learning algorithms is their ability to analyze vast amounts of data and identify patterns and trends that may be difficult for human experts to detect, namely the multidimensional and non-linear nature of musical structure. Furthermore, natural language, although considered from a structuralist perspective as a limited model of something else ([Wittgenstein, 1922](#)), can also be interpreted creatively. Thus, under Deleuze's perspective, language would not be just a system of signs or symbols representing pre-existing concepts or ideas. Rather, language would be a creative force that produces new ways of thinking and forms of expression ([Deleuze & Guattari, 1991](#)).  This means that affective music conditioning by semantics could offer us a new approach to the musical experience.

Using deep learning with natural language has significant implications for music. These tools provide users with a new way to approach their work, opening up new possibilities for exploration and experimentation. They can save users time and effort in generating new ideas and help them focus from another perspective on their work.

## VI. CONCLUSIONS

> Your text prompt                                    Random | Choose
> Generate an uneasy rhythm using a sad wind ensemble and a dark melody.
> ADD MODIFIERS    ADD ANOTHER PROMPT

In this paper, we expose a historical perspective of automatic music generation, from the rules-based expert systems to the deep learning models, to discuss the limits and possibility of expressing affective music conditioned by semantics. Natural language understanding is at the top of human concern, but there are a lot of gaps (and biases) to be solved. So the question of whether words are enough should remain open. But overcoming its limited capacity, language equally leverages musical creativity.

## Footnotes

1. The notion of "expression" does not require a correspondence between what the listener perceives in a piece and what the composer or performer intends to express. In contrast, the concept of "communication" requires that there is both an intention to express a specific emotion and recognition of this same emotion by a listener (Schubert, 2010).

    ↩

2. The KTH rule system models performance principles used by musicians when performing a musical score, within the realm of Western classical, jazz and popular music (Friberg, 2006). ↩

3. https://mubert.com/ ↩

## References

1. Juslin, P., & Västfjäll, D. (2008). Emotional responses to music: The need to consider underlying mechanisms. *BEHAVIORAL AND BRAIN SCIENCES, 31,* 559 –621. ↩
2. Scherer, K. (2000). Psychological models of emotion. *The neuropsychology of emotion,* 137–162. ↩
3. Davies, S. (1994). *Musical Meaning and Expression.* Cornell University Press. ↩
4. Juslin, P., & Laukka, P. (2003). Communication of Emotions in Vocal Expression and Music Performance: Different Channels, Same Code? *Psychological Bulletin, 129(5),* 770-814. 1037/0033-2909.129.5.770 ↩

94.
Deleuze, Gilles & Guattari, Félix (1991). What is Philosophy?. Columbia University Press.

↩